\documentclass[showpacs,amssymb,aps,nofootinbib,floatfix]{revtex4}
\usepackage{epsfig}
\newcommand{\be}{\begin{eqnarray}}
\newcommand{\ee}{\end{eqnarray}}

\newcommand{\ave}[1]{\left\langle #1 \right\rangle}

\newcommand{\eqcomma}{\phantom{AA},\phantom{AA}}

\begin{document}
\title{Quark fluids in heavy ion collisions}
\author{Giorgio Torrieri\\
FIAS,  J.W. Goethe Universit\"at, Frankfurt am Main, Germany,\\ torrieri@fias.uni-frankfurt.de}
%\maketitle
\begin{abstract}
We give a pedagogical introduction (suitable to upper level physics undergraduates) to the field of ultrarelativistic heavy ion collisions.
We pay particular attention to our understanding of the thermodynamic and hydrodynamic properties of the matter created in heavy ion collisions at RHIC energies.
\end{abstract}
\maketitle
Finding the most fundamental blocks of matter has long been considered to be one of the fundamental problems within the physical sciences.

Since a couple of centuries, we have established that the fundamental building blocks of matter as we know it are atoms.    At about the same time, we realized atoms are in fact not fundamental, but are themselves made out of more basic blocks, protons,neutrons and electrons.

Since the seventies, we know that protons and neutrons are made of constituents called quarks and gluons (or sometimes partons).
The quest of better understanding the fundamental components of matter has then hit an obstacle.   Actually, two obstacles.
The first, obvious one, is that we found nothing new since then.  Higher and higher colliding energies  (by the uncertainity principle, the minimal size probed is roughly the inverse of the probing energy) have not yielded any observational result that would indicate that quarks are not pointlike but are themselves composite states.

However, there is a different, more subtle snag:  Since special relativity and quantum mechanics were combined, we know that, because of the equivalence of mass and energy ($E=p^2+m c^2$) and the uncertainity principle ($\Delta E \Delta t \sim \Delta p \Delta x \sim h$), the {\em number of particles} is not fixed but subject to quantum fluctuations.

This introduces a subtelty into the whole concept of ``A is made of B''.
Any system is, potentially, made up of an infinite number of particles, and the contributions of arbitrarily complicated multi-particle systems need to be
included in any calculation (For a technical introduction to the topics 
described here, I recommend \cite{peskin}).

\begin{figure}
\centerline{\epsfig{file=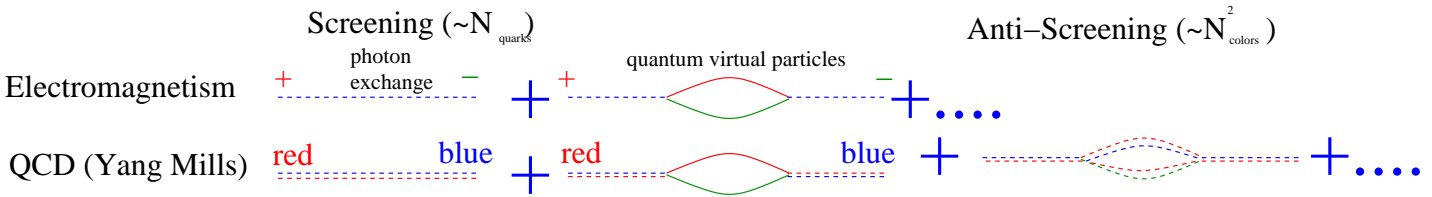,width=19cm} }
\caption{Inter-particle forces in electromagnetism and QCD \label{vacpol}}
\vspace*{-2mm}
\end{figure}

For electromagnetism (or,to be more precise, quantum Electrodynamics) this does not matter much: The quantum many body effects (``vacuum polarization'', shown in Fig. \ref{vacpol} upper panel) reduce themselves to a slight decrease of the effective charge seen by charged particles with distance.  In other words, when the two electric charges come closer together, their binding energy becomes slightly stronger.  If they are {\em really} close, it could become strong enough that ``two charges with a potential'' is not anymore a good description of the system, so many particle-antiparticle pairs will appear from the vacuum   that the series in number of particles will diverge.  However, the distance for this to happen is so short (or the energy so large.  This energy is called the Landau pole) that it simply lies beyond the regime where electrodynamics is a valid theory.

Quarks and gluons, however, interact in a slightly different way, seen 
in the lower panel of the same figure.  It all comes down to the fact 
that in electromagnetism there are only positive particles, negative 
particles, and neutral photons.
Photons can create particle-anti particle pairs, but otherwise do not interact, and the fact that these pairs are {\em massive} means the fluctuations are automatically cut off (the vacuum needs to fluctuate by at least twice the electron mass to create such pairs, and this is unlikely)
%(see Fig. \ref{asympt}).
%\begin{figure}
%\centerline{\epsfig{file=asymptotic.eps,width=15cm} }
%\caption{Asymptotic freedom \label{asympt}}
%\vspace*{-2mm}
%\end{figure}

In QCD, however, quarks can come in three colors and antiquarks in three anti-colors (the way to make a neutral particle is \underline{either} to mix a color with an anticolor, \underline{or} three colors together).  Gluons (the photons of QCD,also massless) carry a color and an anticolor.   In other words, gluons are also charged and can interact with one another, and have more ways of interacting with one another than quarks, since there are 8 gluons (one for each color and different anti-color) and only 3 quarks.   

Most importantly, the effect of these gluon loops on vacuum polarization is {\em opposite} to that of the quarks, {\em anti-screening}:  For small distances charges between quarks are small, for large distances effective charges become larger (gluon vacuum fluctuestions make them increase), and for a distance of about $1 fm$ they diverge:  The system becomes some not yet understood coherent collective system of {\em many} quarks and gluons.
\begin{figure}
\centerline{\epsfig{file=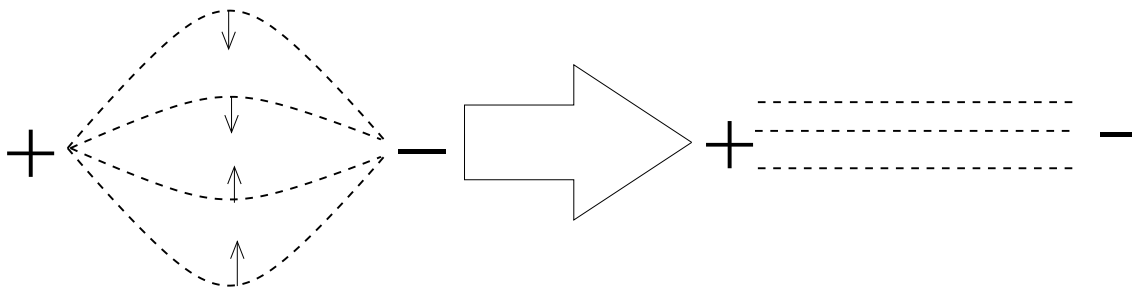,width=15cm} }
\caption{Qualitative confinement \label{confqual}}
\vspace*{-2mm}
\end{figure}
If the original color charges were infinitely heavy, one can intuitively see what happens in Fig. \ref{confqual}:  Field lines interact and (usually) attract.  So if the distance between them is big, they converge to form a tube.  And therefore, by Gauss's law the force between two charges stops decreasing with distance, since the area of the tube remains constant.  In fact, due to the anti-screening effect it increases.
Since most quarks are {\em not} heavy, but actually $\sim$ the electron mass, the tube can only be stretched until a pretty small ($\sim 1$ fm, $10^{-15}$ m) distance.  Beyond it, the energy field will become strong enough to make new {\em stable} quarks out of the vacuum, and the string ``breaks'' into new color-neutral quark-antiquark pairs.   The mass of protons is, to an acceptable approximation, nothing else but the uncertainity principle estimate from this size, $h/(1fm*c), \sim  1 GeV$.  Below these distances, strings are short enough that they ``stretch out'', and charge is small enough that perturbation theory applies.

\begin{figure}
\centerline{\epsfig{file=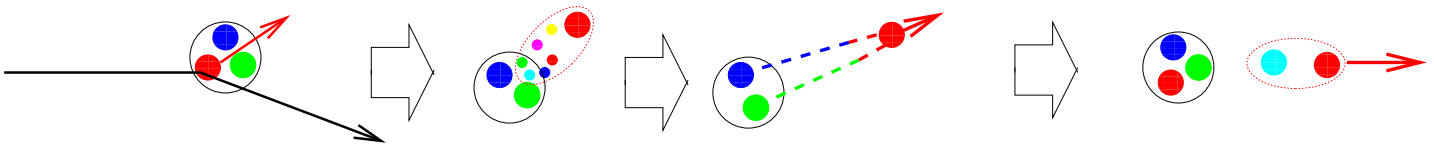,width=19cm} }
\caption{Deep inelastic scattering \label{dis}}
\vspace*{-2mm}
\end{figure}

The above reasoning suggests that a good description of strongly interacting particles (called ``hadrons'') is 
that of ``bags'' of free quarks:  As long as the quark stays inside the bag, it behaves as a free particle (no room for strings).    But if it is kicked out of the bag, it develops a string, that ends up breaking into quark-antiquark pairs which end up forming color neutral objects, hadrons moving close together.  As long as we do not worry about the details these hadrons (treat each correlated bunch of fast hadrons as a ``jet'' and assume its energy is that of the original quark) this picture allows us to calculate cross-sections for deep-inelastic scattering $e-p$ collision experiments \cite{deepine} Fig \ref{dis} which have allowed us to test weakly coupled QCD thoroughly.   The theory works to the extent we are sure it is true.  We just can not see the fundamental degrees of freedom directly.

So, would it be possible to produce ``free quarks'' that can travel and interact over a large volume?   The previous discussion does leave a way out:
If the distance between a bunch of protons and neutrons becomes comparable to the proton size, the bags ``melt'' and quarks behave as normal particles, interacting via a weak potential.

   One way to do this would be to compress ``a lot of'' nuclear matter 
to the point where the average separation between the nuclei would be 
less than the size of the proton.   Quarks would then be free to jump 
from bag to 
bag, so the bags would, presumably simply disappear.   Another way to see it is to realize that the such a compressed system becomes so hot that quarks will be able to overcome the string-like potential and travel from quark to quark.  Such a picture is strongly suggestive of a phase transition, similar to the phase transition between ice and water (which occurs when molecules get enough average energy to escape the intermolecular forces which keep them in their position in the ice crystal).

What is the phase transition temperature from the ``hadron phase'' to the ``quark phase''?
 A quantitative estimate for this is to compare the proton size ($1 fm$) with the density of an ideal gas of 4 types of Fermions (up and down quarks,spin up and down) and 16 types of gluons(8 and 2 polarizations).  The density is (In natural units where $\hbar=c=1$ and $h c = 0.197 GeV fm$) \cite{ll2}
\begin{equation}
\label{density}
n =  \frac{2 \pi^2}{180} \left( \frac{7}{8} N_{quarks} + N_{gluons}  \right) T^3 \simeq 2.136 T^3
\end{equation}
If for quark liberation the mean distance between quarks, $n^{1/3} \sim 1 fm$
it leads to a temperature of 
$\simeq 152$ MeV (a trillion or so degrees).   More sophisticated models 
based on lattice QCD \cite{lattice} put this temperature to be a surprisingly similar 150-200 MeV, leading to the suggestion that soon after the phase transition quarks become a nearly ideal gas describeable by perturbation theory techniques.   A millionth of a second after the big 
bang, the whole universe was such a gas of quarks and gluons, called Quark GLuon Plasma (QGP).

We do not, however, know much more about this state of matter.  Is it really weakly coupled or do strong interactions persist?  How does the transition from quark gluon matter to normal matter proceed?  Is it {\em really} a phase transition, like ice and water, or is it a cross-over, like atoms and electromagnetic plasma?   

\begin{figure}
\centerline{\epsfig{file=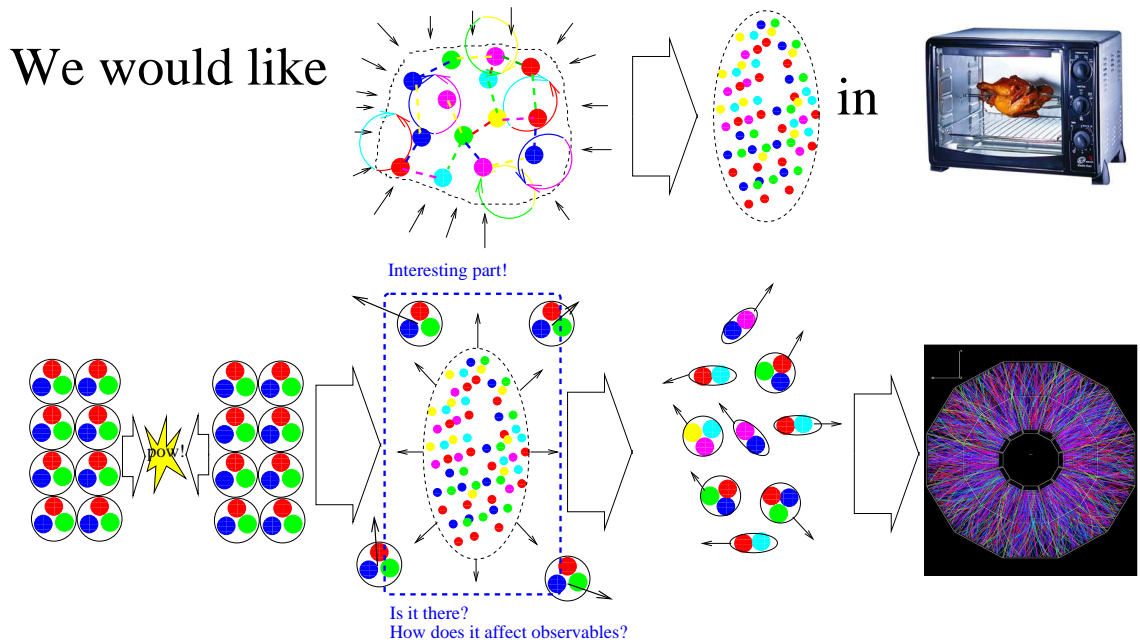,width=15cm} }
\caption{What we need to study thermodynamics and what we get in heavy ion collisions \label{collision}}
\vspace*{-2mm}
\end{figure}

To explore these questions empirically is to create a chunk of quark-gluon matter in the laboratory.    Fig. \ref{collision} shows the only way we can think of to do it:  We collide two ``large'' nuclei at very high energy, and hope that some of this energy goes into heat.  Enough heat to create a locally equilibrated liquid of quark matter, whose thermodynamic properties we then study.

The trouble is that this is {\em not} the same as heating up matter in an ``infinitely slowly heating oven'' where thermodynamics applies automatically (Fig \ref{collision}).
  Even if the quark matter was created, it will soon expand, transform into hadrons, and decouple.   Finding an unambiguous experimental sign that quark matter was in fact created is {\em not} a trivial matter.   In fact, confirming experimentally quark matter was created is a still unresolved question.

A plethora of experimental probes have been proposed to resolve it, ranging from quark chemistry (strangeness enhancement), to elecromagnetic signals (photons, dileptons), to the study of fluctuations (In context of susceptibilities diverging in phase transitions), to the behaviour of very high energy particles (``jets'') and heavy quark bound states. The interested reader can find a good introduction to these in \cite{foundations,jansbook} and the current experimental situation in \cite{BRAHMS,PHENIX,PHOBOS,STAR,CERN}.  In a sentence, we do not know enough and giving a balanced picture would take many more pages than allowed here.   
So I will devote the remainder of this work to explore an even a more basic question:

  Phase transitions and thermodynamic properties are relevant for systems in local thermal and chemical equilibrium.
It is not at all clear that a short-lived, fast expanding system is in local equilibrium to the extent that thermodynamic concepts such as phase transition and temperature will be relevant at all.

\begin{figure}
\centerline{\epsfig{file=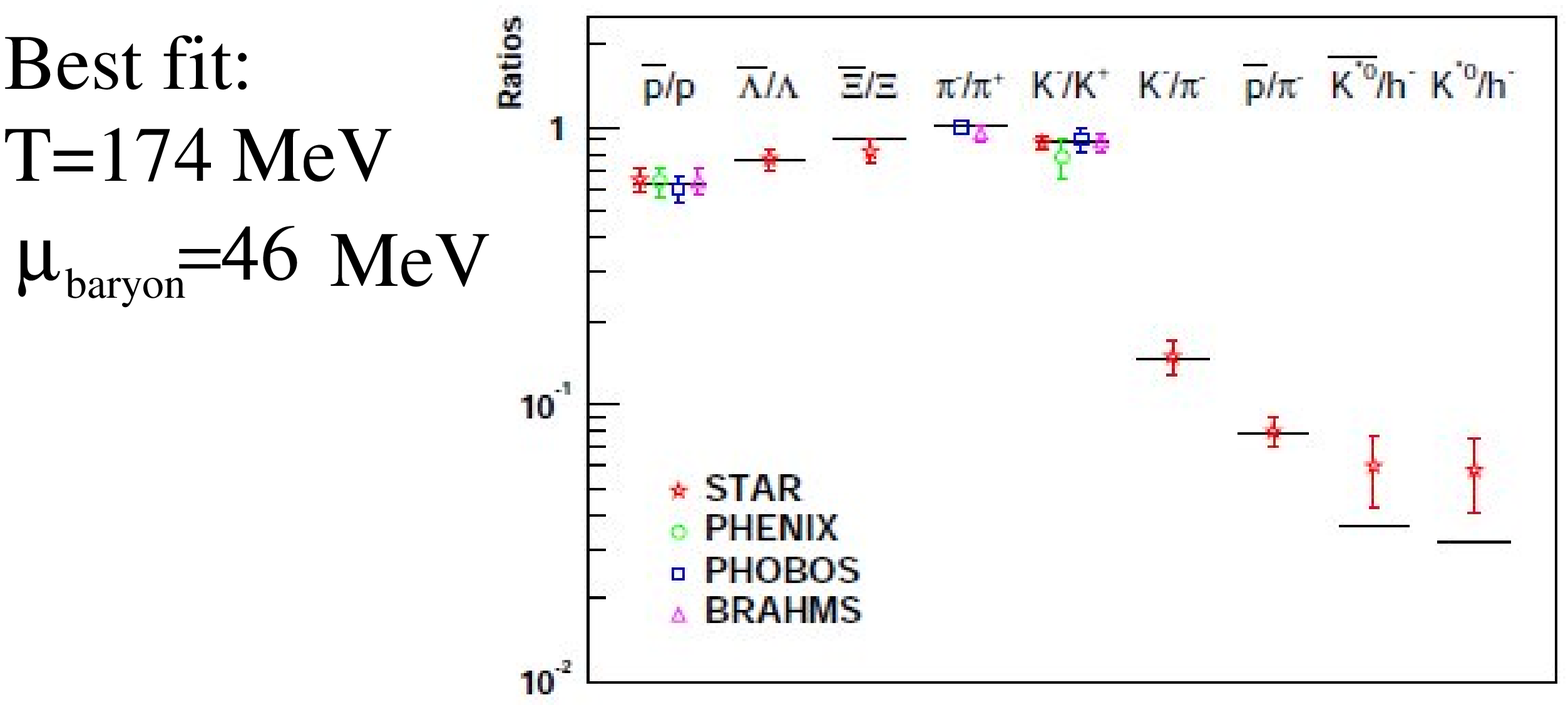,width=15cm} }
\caption{Thermal particle fit to particle ratios  at RHIC \cite{BRAHMS,PHENIX,PHOBOS,STAR} \label{kanetafig}}
\vspace*{-2mm}
\end{figure}

In fact, the idea to describe strong interactions by statistical mechanics techniques, where the final abundance of particles obeys some kind of thermal distribution in terms of the particle's degeneracy $g^*$ and temperature (taking into account feed-down from resonances decays with branching ratios  $b_{j \rightarrow i}$)
\begin{eqnarray}
n^{thermal}_i \sim g^*_i \int \frac{d^3 p}{1 \pm \exp \left[ \frac{\sqrt{p^2+m_i^2}-\mu}{T} \right]}\eqcomma n^{final}_i = n^{thermal}_i + b_{j \rightarrow i} n^{thermal}_j
\end{eqnarray}
This thermodynamic approach has a long and illustrious history ~\cite{Fer50,Pom51,Lan53,Hag65}.  It was using these techniques that \cite{Hag65} has managed to predict that hadronic gas had a phase transition even before the existance of quarks \footnote{The reasoningwent as follows \cite{Hag65}: it was noted that the statistical degeneracy of strongly interacting particles increases with its mass, so not only these particles can have arbitrary mass, the more massive particles have greater statistical weight.  It was reasoned that for heavy enough particles, distinguishing a ``particle'' from a ``thermalized fireball'' will be impossible, so we might as well consider all particles to be blobs of hot matter (``fireballs are made of fireballs'').  This constrains the partitiion function to obey the recursion relation
\[\
 Z(V,T) =
\int d^3 p dm \rho(m) \exp\left[ \frac{-\sqrt{p^2+m^2}}{T} \right] \Rightarrow Z(V,E) = dE' \int Z(V,E') \exp\left[ \frac{-E'}{T} \right] dE'  \]
It can be shown that this forces the density of states $\rho_m$ to be exponential, $\rho(m) \sim \exp \left[ \beta_H m \right]$
Hence, at $T>\beta_H^{-1}$ the partition function diverges
}
Today, it is very well known that a fitted temperature, chemical potential and volume can describe the abundances of all strongly interacting particles produced in heavy ion collisions \cite{jansbook,bdm} (Fig. \ref{kanetafig} for the simplest model, although extensions are possible, see \cite{jansbook})  The temperature fitted is very close to the phase transtion temperature at highest energies, and a universal freeze-out condition such as Energy/Baryon $\simeq 1$ at lower energies.   

The interpretation of this finding is, however, hotly contested, with 
some people attributing it to more mundane processes (in multi-particle 
collisions phase space weights dominate over dynamics, and that gives 
you approximately statistical mechanics \cite{Fer50}) to very exotic 
ideas (strongly interacting particles are ``born in equilibrium'' 
\cite{castorina}, like 
Hawking radiation \cite{hawking} from a black hole).   
Related 
to this discussion is the controversy of {\em how small } does system have to be before statistical models are valid.   Qualitatitatively, even $e^+ e^- \rightarrow hadrons$ looks statistical  ~\cite{Fer50,Pom51,Lan53}.   Quantitatively, different groups give different conclusions as to how good the fit it \cite{castorina,pbmel}
\begin{figure}
\centerline{\epsfig{file=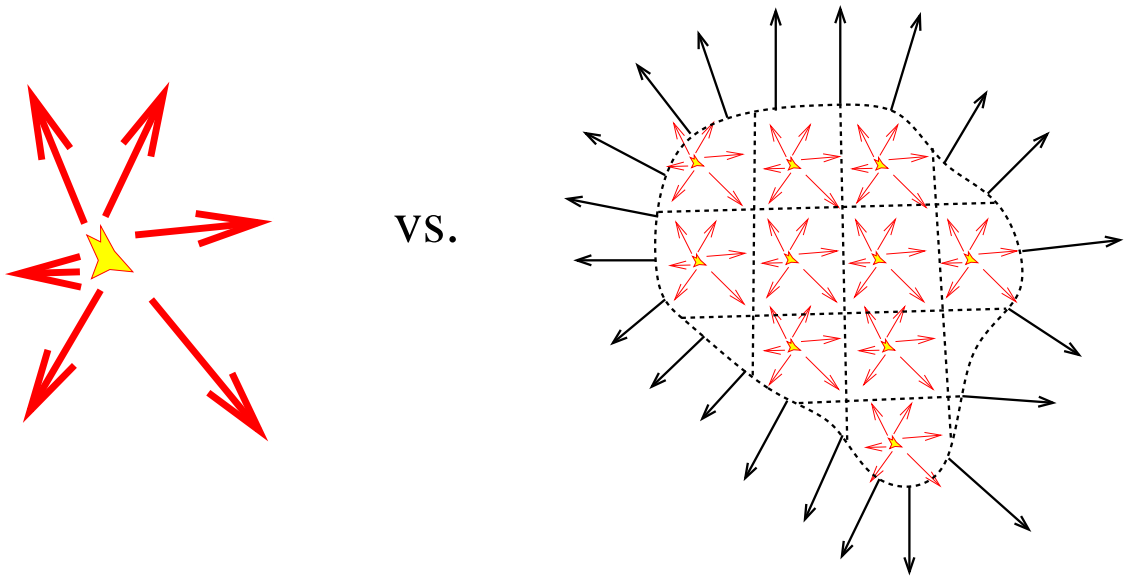,width=15cm} }
\caption{Global vs local equilibrium \label{glocal}}
\vspace*{-2mm}
\end{figure}
Here we evade this controversy by requiring that matter is not just ``globally in equilibrium'' (particles look like they have a temperature) but ``locally in equilibrium'' (particles have a temperature, pressure and density {\em at each point}.   See Fig. \ref{glocal} for the difference.  This, invariably, leads us to consider not just temperature, but hydrodynamic flow.  In fact, flow can be used as a tool to ascertain the thermodynamic and hydrodynamic properties of the system.

Quantitatively, a locally equilibrated ``fluid'' can be distinguished from a far-from equilibrium ``dust'' of particles by using the Knudsen number \cite{ll1}, $K= l_{mfp}/L$ where $l_{mfp}$ is  the mean free path between two ``microscopic'' collisions and $L$ the ``macroscopic'' system size.
It is clear that when $K \ll 1$, each particle in the system interacts ``many many times'' with the other particles, and the system is, to a good approximation, a fluid where locally thermal equilibrium has been achieved.  If, on the other hand, $K \geq 1$ the system is a ``dust'', where particles to a good approximation ignore each other and statistical mechanics concepts such as temperature are inapplicable.

As we know from \cite{ll1}, the mean free path for a weakly coupled gas is $1/(\sigma n)$ where $\sigma$ the scattering cross-section between two particles and $n$ the particle density, which for the massless gas limit is $s = 4 n$ where $s$ is the entropy density.   In turn, in leading order perturbation theory $\sigma$ is related to the average momentum as $\sigma \sim \frac{g^4}{\ave{p}^2}$, and by the massless gas formula $\ave{p} = 3 T$.   Putting everything together, we get the following estimate for the Knudsen number \cite{ll1,ll2} and Eq. \ref{density} 
\begin{equation}
K_{weak} = \frac{4}{ \sigma s L} = \frac{36}{8.144 g^4 L T}  
\end{equation}
remembering that for deconfinement $T=200$ MeV, for a nucleus $L \sim 10$ fm, and for perturbation theory to be valid $g \ll 4 \pi$ (say $\sim 1$ stretches it), we get a Knudsen number of $0.4$, out of equilibrium by $40 \%$.  Hence, our hope of the system locally equilibrating are not really justified if the system is really perturbative.

What if the system is {\em not} perturbative?  Well, the very short answer is we dont know how to calculate things than, but we naively expect viscosity to be small since the system is hot and strongly coupled (lots of collisions, and ``on average'' each collision brings the system closer to equilibrium by Boltzmann's H-theorem \cite{ll1})

Could we do better than that?  Well, if $g \gg 1$ we can not anymore use perturbation theory.   From quantum mechanics, however, we can guess that the mean free path can not be smaller than the De Broglie wavelength \cite{minvisc} (if it could, a thin layer of this material could function as a detector capable of breaking the uncertainity principle).   Hence, provided $\ave{p}\sim 3 T$ is still valid the Knudsen number would be something like
\begin{equation}
K_{strong} \sim \frac{1}{\ave{p}L} \sim \frac{1}{3 T L} \sim 0.01  
\end{equation}
so the system would be very nearly thermalized (even for $p-p$ collisions $K\sim 0.3$)  Of course, the $\ave{p}\sim 3 T$ assumption is completely unjustified in such a strongly coupled system, and the mean free path is not well defined since the number of particles will be extremely ill-defined.
So this is at best a {\em very} rough estimate.

Can we do any better?   Recently, it was discovered that a theory ``similar to QCD'' (QCD with a large number of colors and 4 supersymmetries)  is ``dual'' to a string theory defined on a 10 dimensional space with a string coupling inverse to the field theory coupling constant \cite{maldacena}   (Note that such a theory is ``conformal'', so the coupling constant does not run.  This conjecture, and it has not been proven yet, is known as the AdS/CFT correspondence).   

This means that an ``infinitely strongly coupled theory'' with infinite 
number of colors is equivalent to a weakly coupled string theory defined 
on a 5-dimensional space.    The latter is simply classical general 
relativity and Hawking's formula for black hole entropy \cite{hawking}.
``Equivalent'' simply means that it is possible to define a dictionary to convert any Gauge-invariant observable in the quantum field theory into a corresponding observable in the string theory.  The latter can be calculated since this theory is weakly coupled, and the answers will ``magically'' be the same.   

In particular, the entropy density becomes the ``area'' of a 5-dimensional black hole (calculable from Hawking formula), and the viscosity becomes the speed at which small perturbations of the black hole dissipate (calculable in general relativity).   This led to the ``famous'' entropy and viscosity formula  for an ``infinitely strongly coupled theory'' (of a very particular type) \cite{kss}
$\frac{\eta}{s} = \frac{1}{4 \pi}$.
Plugged into our previous Knudsen number formulae this gives again a Knudsen number of 0.01 or so.   So the system should certainly be a perfect fluid in the strongly coupled limit.

\begin{figure}[t]
\centerline{\epsfig{file=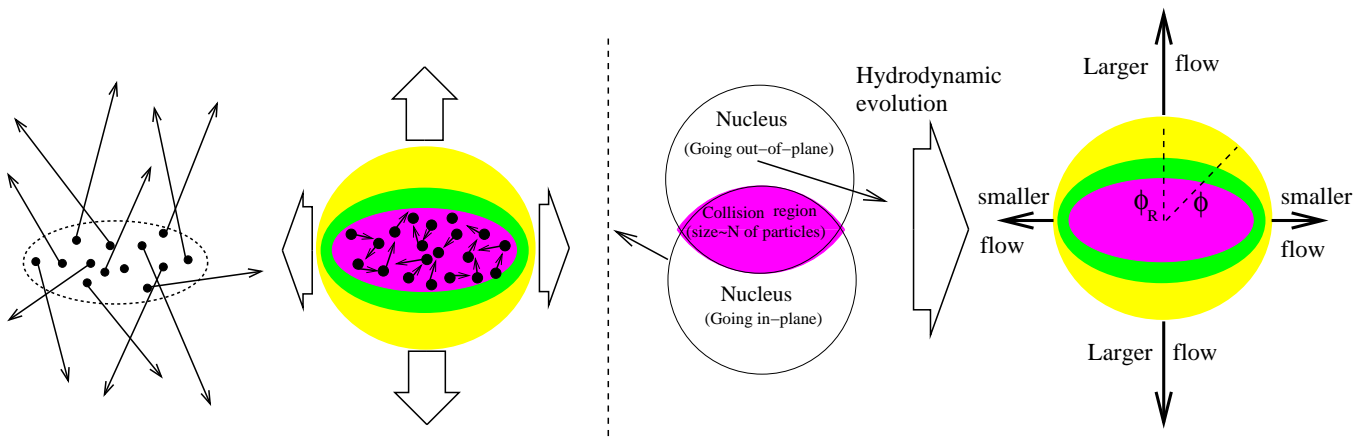,width=18cm} }
\caption{Left panel: A dust (particles move independently of the initial
shape) vs a fluid (particles move under forces determined by the initial
density gradients).  Right panel: The expected expansion of a fluid
originating from an ellipsoidal hot region in a non-central collision.
\label{fluid}}
\vspace*{-2mm}
\end{figure}

Can all these ideas give experimentally measureable consequences?   Well, we know what effect viscosity has on a fluid: It acts as a friction, transforming gradients of flow into heat \cite{ll1}.     Hence, a ``perfect fluid'''s final momentum is determined only by initial gradients in density.   In contrast, in a viscous fluid particles dont interact with each other, so gradients in density have no effect on the final kinematics.   This can be see in FIg.  \ref{fluid}.   Viscous hydrodynamics should interpolate between these two extreme cases.

The assumption of perfect thermal equilibrium at every instant is equivalent to assuming that in {\em some} frame the Energy-momentum tensor is diagonal, and pressure is isotropic (the same in all directions), so \cite{ll1}
\begin{equation}
T_{\mu \nu}^{comoving} = \left( 
\begin{array}{cccc}
e & 0 & 0 &0 \\
0 & p & 0 &0\\
0 & 0 & p &0\\
0 & 0 & 0 &p
\end{array}
   \right) \eqcomma T_{general}^{\mu \nu} = e u^\mu u^\nu + p \Delta^{\mu \nu}
\end{equation}
where the projector $\Delta^{\mu \nu} = u^\mu u^\nu - g^{\mu \nu}$ projects a rank-2 tensor in the direction perpendicular to the flow $u_{\mu}$. Here  $g^{\mu \nu}$ is the metric tensor, $p$ the pressure and $e$ the energy density.  The latter are related by statistical mechanics (for the ideal massless gas discussed earlier $p=e/3$).   This means that, for a perfect fluid, the conservation equation 
$\partial_\nu T^{\mu \nu} =0$
becomes solvable from any set of initial conditions,since there are 5 equations (4 conservation equations and an equation of state) and 5 unknowns ($u_{x,y,z},p,e$).     These can be considered as the equations of motion of a collective system to ``0-th order'' in the Knudsen number, assuming the ``macroscopic scale'' is given by the flow gradient $\partial u$.

For viscous hydrodynamics, we have the additional equations to define shear and bulk viscosity $\eta,\zeta$ in terms of temperature and the energy momentum tensor is more complicated
\begin{equation}
T^{\mu \nu} = \left. T^{\mu \nu}\right|_{equilibrium} + \Pi_{\mu \nu}
\end{equation}
\begin{equation}
\Pi_{\mu \nu} = -\eta \left(  \Delta_{\alpha \mu} \partial^\alpha u_\nu + \Delta_{\alpha \nu} \partial^\alpha u_\mu - \frac{2}{3} \Delta_{\alpha \beta} \Delta^{\alpha \gamma} \partial^\beta u^\mu \right) - \zeta \partial_\alpha u^\alpha \Delta^{\mu \nu}
\end{equation}
but still solvable provided we can calculate $\eta,\zeta$ in terms of the temperature.  Since $\eta,\zeta \sim l_{mfp} \ave{p} n$ \cite{ll1} these can be considered the equations of motion to first order in the Knudsen number\footnote{In fact, these equations have problems with relativstic causality, and have to be extended with terms to {\em second order} in the Knudsen number.  See \cite{romatschke} for more details on this}
(The complicated looking dissipative tensor $\Pi_{\mu \nu}$ is simply the traceless flow-perpendicular part of the gradient of the flow).
Note that these equations are non-linear, so in general they can not be solved analytically but just integrated numerically from a set of initial conditions.

The physical problem to which this formalism is to be applied is shown again in the right panel of Fig. \ref{fluid}.    Initially, the system in a typical non-central collision, is, approximately, an ellipse, so its density transverse to the beam is
\begin{equation}
e(r,\phi) = e_0(r) \left( 1 + 2 \epsilon \cos 2 (\phi - \phi_R) \right)
\end{equation}
where $\phi_R$ is the reaction plane, seen in Fig. \ref{fluid}

Integrating the equations shown earlier with these initial conditions \footnote{Initial conditions also need to give the density profile {\em parallel} the the beam. This is usually taken to be ``Boost-invariant"', essentially assuming that the fluid flows apart in the beam direction with a Hubble type initial velocity $v_z = z/t$.  See \cite{bjorken} for a description justification of this initial condition.  Note that even in the opposite (``Landau'') limit \cite{Lan53} where there is no longitudinal flow and nuclei ``stick together'', a Bjorken-like flow develops very quickly, after a few fm. The applicability of these conditions, and the effect of them on the results, is currently not very well undestood } is involved and can not be done analytically.  However, it can be readily seen that, if viscosity is small, the initial gradient in density will transform into a gradient in transverse flow $v_T$
\begin{equation}
u_T (r,\phi) = u_{T0} \left( 1 + 2 \epsilon_v \cos 2( \phi - \phi_R) \right)
\end{equation}
At the moment, we do not know how this fluid breaks up into particles.
However, this break-up has to conserve momentum locally.   This means that the average transverse momentum, as a function of $\phi$ will also be something like
\begin{equation}
\ave{p_T}(\phi) = \ave{p_T}_0 \left( 1 + 2 v_2 \cos 2( \phi - \phi_R)  \right)
\end{equation}
and this leads to a Fourier coefficient in the average $p_T$ binned by $(\phi- \phi_R)$.  This coefficient is known as $v_2$ \cite{v2orig}.
$v_2$ is calculable within hydrodynamics, measurable by experiments, and highly sensitive to viscosity during the initial stage (Since hydrodynamic forces tend to make the fireball rounder, it is thought that $v_2$ stops forming when the system expands \cite{v2orig}.  Note that the role of $v_2$ in the final stages of the fireball evolution is currently controversial). 

\begin{figure}[t]
\centerline{\epsfig{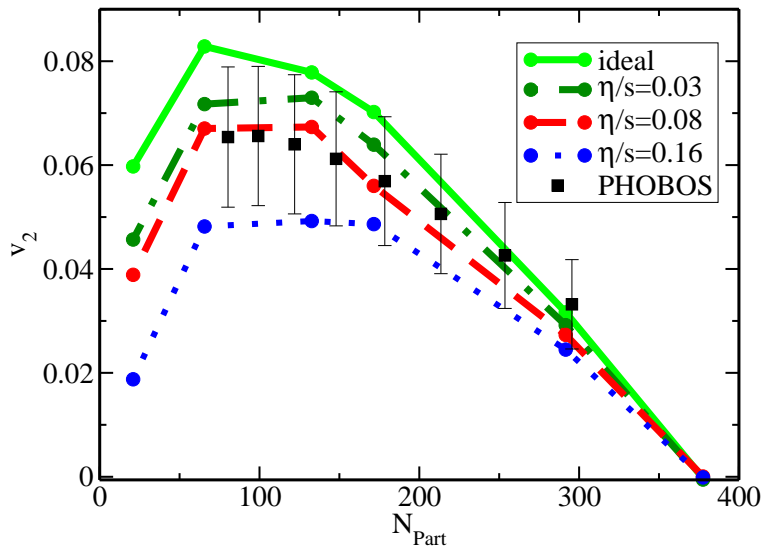} }
\caption{Experimental results from $v_2$ modelling of hydrodynamics,
from \cite{romatschke}.  $N_{part}$, the number of participants,
measures the overlap area of the collision, see previous figure
\label{romatschke}}
\vspace*{-2mm}
\end{figure}

Comparing theoretical calculations to experimental measurments of $v_2$ 
\cite{BRAHMS,PHENIX,PHOBOS,STAR}, it can be seen that the system created at RHIC appears to be a very good fluid, well in the strongly coupled regime described earlier (Fig \ref{romatschke}) \cite{sqgpmiklos,sqgpshuryak,romatschke}.  
In fact, viscosity can not be too much bigger than $\eta \sim s/4\pi$, the lower bound value calculated through the AdS/CFT correspondence \cite{kss}.
This has made a lot of people excited at the prospect that string theory can be used to describe the system created in heavy ion collisions.

In this work  we have concentrated on proving that a collective locally thermalized good fluid is indeed created in heavy ion collisions.   What we did not say much about are the properties of this fluid:  Is it really made of quarks and gluons?  What is the nature of the transition between this fluid and normal hadronic matter?  
At what center of mass collision energy does such a ``perfect fluid'' form?  How big does it really need to be before these collective effects arise?  At what temperature, if any, does the system become weakly coupled and perturbation theory can be used?

Other than lack of space, the main reason I did not mention these topics is that there is not much consensus about them.   However, a lot of experimental activity is planned to try and clarify these questions.  The focus is to vary as much as possible energy,system size, type of nucleus, collision geometry etc, to see if ``something interesting'' (eg a jump in $v_2$, or multiplicity, or whatever) happens in some particular regime.

The recently opened LHC accellerator plans to conduct very high energy heavy ion experiments, where we hope the temperature will be much bigger than the critical temperature. Further into the future, experiments with {\em lower} energy are planned to determine the critical threshold for production of quark matter.
Two experiments of this type with a lot of Russian involvement are planned at FAIR in Germany \cite{fair} and NICA in Dubna \cite{nica}.

In conclusion, I heartily recommend Baikal school students to explore this subject further.   It can be very theoretical yet includes heavy contact with experimental data.  It draws insights from fields as far apart as thermodynamics and string theory.   And, above all, it is very far from being thoroughly, or even approximately understood.   In other words, we need your help!

G.T. acknowledges the financial support received from the Helmholtz International
Center for FAIR within the framework of the LOEWE program
(Landesoffensive zur Entwicklung Wissenschaftlich-\"Okonomischer
Exzellenz) launched by the State of Hesse.
GT thanks the organizing committee for the Baikal school for theoretical physics for inviting me to this great school, and all participants for questions, comments and discussions.

\end{document}